# Large thermoelectric power factors in black phosphorus and phosphorene

*Hongyan Lv,[1] Wenjian Lu,[1,*] Dingfu Shao,[1] and Yuping Sun[1,2,3,*]*

[1] Key Laboratory of Materials Physics, Institute of Solid State Physics, Chinese Academy of Sciences, Hefei 230031, People's Republic of China

[2] High Magnetic Field Laboratory, Chinese Academy of Sciences, Hefei 230031, People's Republic of China

[3] University of Science and Technology of China, Hefei 230026, People's Republic of China

**Corresponding Author**

**\*** Email: wjlu@issp.ac.cn. **\*** Email: ypsun@issp.ac.cn

The electronic properties of the layered black phosphorus (black-P) and its monolayer counterpart phosphorene are investigated by using the first-principles calculations based on the density functional theory (DFT). The room-temperature electronic transport coefficients are evaluated within the semi-classical Boltzmann theory. The electrical conductivity exhibits anisotropic behavior while the Seebeck coefficient is almost isotropic. At the optimal doping level and room temperature, bulk black-P and phosphorene are found to have large thermoelectric power factors of 118.4 and 138.9 $\mu Wcm^{-1}K^{-2}$, respectively. The maximum dimensionless figure of merit ($ZT$ value) of 0.22 can be achieved in bulk black-P by appropriate $n$-type doping, primarily limited by the reducible lattice thermal conductivity. For the phosphorene, the $ZT$ value can reach 0.30 conservatively estimated by using the bulk lattice thermal conductivity. Our results suggest that both bulk black-P and phosphorene are potentially promising thermoelectric materials.

**TOC GRAPHICS**

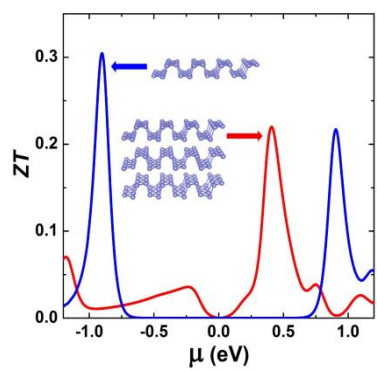

Thermoelectric materials, which can directly convert heat into electricity and vice versa, have attracted much interest from the science community due to the current critical energy and environmental issues. The performance of a thermoelectric material is quantified by the dimensionless figure of merit $ZT = S^2 \sigma T / (\kappa_e + \kappa_p)$, where $S$ is the Seebeck coefficient, $\sigma$ is the electrical conductivity, $T$ is the absolute temperature, and $\kappa_e$ and $\kappa_p$ are the electronic and lattice thermal conductivity, respectively. It is challenging to achieve a high $ZT$ value since optimizing one transport coefficient often leads to another adversely affected. Recently the break-through works which achieve significant improvement in thermoelectric efficiency have focused on the multi-component systems, such as $Bi_2Te_3/Sb_2Te_3$ superlattice,[1] nanostructured BiSbTe bulk alloys,[2] $AgPb_mSbTe_{2+m}$,[3] $PbTe_{1-x}Se_x$,[4] Na doped PbTe-SrTe all-scale hierarchical architectures.[5] Apart from the complicated fabrication technologies used to synthesize these compounds, most of them contain the elements which are either expensive or toxic.

As for the single-component materials which are constituted by the naturally abundant and nontoxic elements, several groups have investigated the possibility of using carbon and silicon systems for thermoelectric applications. For the carbon based materials, both the pristine graphite and graphene are found to have extremely high thermal conductivity (~2000 $Wm^{-1}K^{-1}$ for graphite[6] and suspended single-layer graphene[7]) at room temperature. As a result, neither graphite nor graphene is considered a potential candidate for thermoelectric applications. Many efforts have been made to enhance the thermoelectric performance of the carbon based materials. For example, Sevinçli *et. al*[8] theoretically investigated the effect of the edge disorder on the thermoelectric properties of the zigzag graphene nanoribbons and found that the phonon thermal conductivity could be reduced while keeping the electronic conductance weakly affected in the edge-disordered nanoribbons, hence the $ZT$ value could be largely enhanced. As for another elemental material, bulk silicon has a considerable Seebeck coefficient; however, due to its large lattice thermal conductivity (~150 $Wm^{-1}K^{-1}$), bulk silicon is a poor thermoelectric material ($ZT$ ~0.01 at 300 K)[9] as well. Efforts mainly focused on reducing the thermal conductivity of the silicon systems, such as by synthesizing the silicon nanowires[10] and holey silicon structures.[11] Although improvements of the thermoelectric performance could be made, the synthesis of the above high quality low-dimensional structures remains a challenge.

Black phosphorus (black-P) is another elemental solid which is the most stable form among the phosphorus allotropes[12] under normal condition. Similar to graphite, black-P crystallizes in a layered structure, namely, each phosphorus atom is covalently connected to three neighboring phosphorus atoms to form a puckered layer. Between the adjacent layers, the interactions are mainly of the van der Waals type. Different from graphite and bulk silicon, the bulk black-P exhibits much lower thermal conductivity, that is, 12.1 $Wm^{-1}K^{-1}$ at room temperature;[13] on the other hand, the bulk black-P is a narrow-gap semiconductor with the band gap of about 0.35 eV.[14,15] Such two factors are very beneficial to the thermoelectric application. Very recently, few-layer and even monolayer black-P (which is called phosphorene) have been successfully exfoliated from the bulk.[16,17,18] However, researches on the thermoelectric properties of bulk black-P and phosphorene are very limited. Only a few previous works reported on the electrical conductivity[14,15,19,20] and thermal conductivity[13] of this system. It is of interest to explore the possibility of using black-P and phosphorene as thermoelectric materials.

In this work, we investigate the structural, electronic, and thermoelectric properties of the bulk black-P and phosphorene. We firstly study the structural and electronic properties by using the first-principles calculations. The electronic transport coefficients at room temperature are then investigated based on the semi-classical Boltzmann theory. By using the experimentally measured lattice thermal conductivity, we evaluate the thermoelectric performance of the bulk black-P and phosphorene. Our calculations indicate that both of them are very promising thermoelectric materials.

We begin our discussion with the layered structure of bulk black-P. The side- and top-views are demonstrated in Figs. 1(a) and (b), respectively. The crystal structure of bulk black-P is orthorhombic with the space group *Cmca*, as shown in Fig. 1(c). The primitive cell contains four

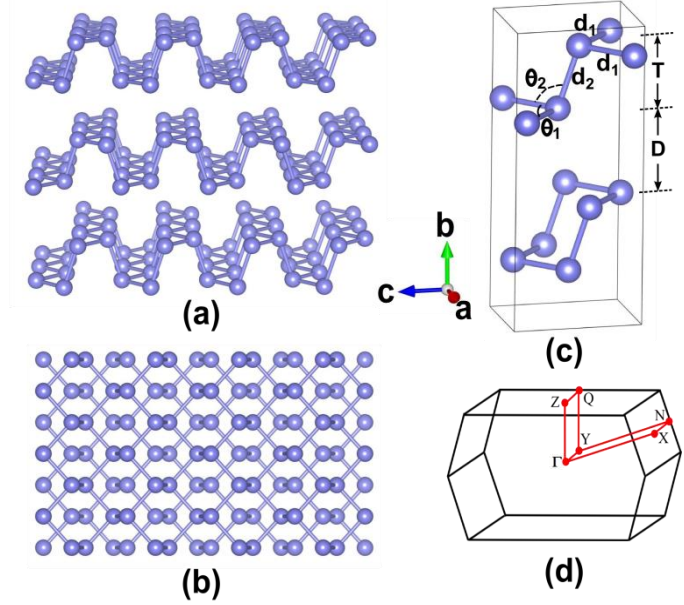

**Figure 1** (a) Side- and (b) top-views of the layered structure of bulk black-P with (c) the crystal structure and (d) the first Brillouin zone corresponding to the primitive cell.

atoms; the corresponding first Brillouin zone is illustrated in Fig. 1(d). The phosphorene is cleaved from the bulk structure and the geometry and atomic positions are fully relaxed. In Table I, we summarize the structural parameters as denoted in the crystal structure of bulk black-P (see Fig. 1(c)) and the corresponding values for phosphorene. The lattice parameters of the bulk black-P are $a$=3.34 Å, $b$=10.51 Å and $c$=4.43 Å, which are very close to the experimental values.[21] When the phosphorene is cleaved from the bulk black-P, the bond lengths $d_1$ and $d_2$ remain unchanged, while the $c$ vector is 4.5% enlarged. As a result, the bond angle $\vartheta_2$ between the bonds $d_1$ and $d_2$ is increased by 1.8% and the phosphorene becomes thinner, with the thickness $T$ 1.4% smaller than that in the bulk black-P.

**Table I** Structural parameters of bulk black-P and phosphorene.

|  | Reference | $a$ | $b$ | $c$ | $d_1$ | $d_2$ | $\vartheta_1$ | $\vartheta_2$ | $D$ | $T$ |
|---|---|---|---|---|---|---|---|---|---|---|
| Bulk black-P | Experiment[21] | 3.31 | 10.47 | 4.37 | 2.22 | 2.28 | 96.5 | 101.9 | 3.08 | 2.16 |
|  | This work | 3.34 | 10.51 | 4.43 | 2.24 | 2.28 | 96.2 | 102.1 | 3.09 | 2.16 |
| Phosphorene | This work | 3.32 |  | 4.63 | 2.24 | 2.28 | 95.8 | 103.9 |  | 2.13 |

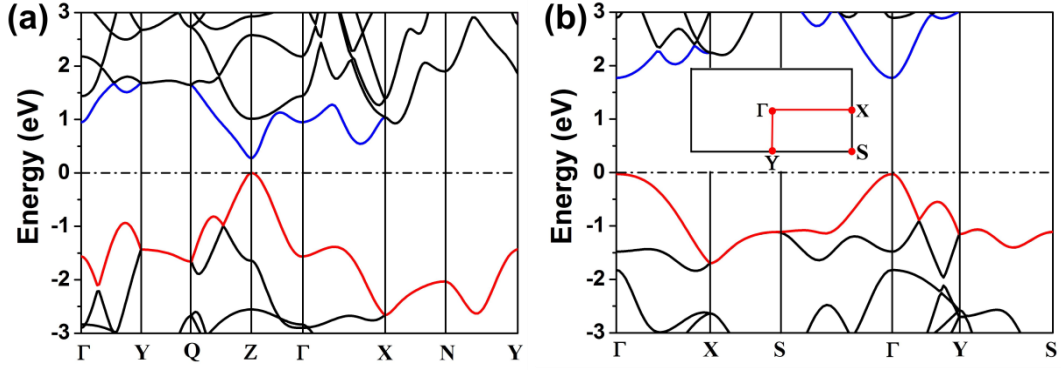

**Figure 2** Band structures of (a) bulk black-P and (b) phosphorene, with the first Brillouin zone used for phosphorene inserted.

Figure 2(a) shows the calculated band structure of bulk black-P with respect to the primitive cell. The red and blue lines represent the bands near the valence band maximum (VBM) and conduction band minimum (CBM), respectively. The result shows that bulk black-P is semiconducting with a direct band gap of 0.34 eV at the $Z$ point, in good agreement with those reported experimentally.[14,15] Our calculation confirms that the TB-mBJ potential can accurately predict the band gap of our investigated system, which is an important factor in determining the electronic transport properties. The band structure of the phosphorene is shown in Fig. 2(b), with the first Brillouin zone inserted in the figure. One can note that phosphorene is a direct-band-gap semiconductor as well. However, due to the quantum confinement effect, the phosphorene has a much larger band gap, 1.80 eV, located at the $\Gamma$ point.

Based on the calculated band structure, the electronic transport coefficients of the bulk black-P and phosphorene can be evaluated by using the semi-classical Boltzmann theory and rigid band model. To get reliable results, a very dense $k$ meshes up to 2954 and 840 points in the irreducible Brillouin zone (IBZ) were used for the bulk black-P and phosphorene, respectively. Within this method, the Seebeck coefficient $S$ can be calculated independent of the relaxation time $\tau$; however, the electrical conductivity $\sigma$ is calculated with $\tau$ inserted as a parameter, that is, what we obtain is $\sigma/\tau$. The relaxation time can be determined by comparing the calculated $\sigma/\tau$ and the electrical conductivity $\sigma$ reported experimentally at a particular carrier concentration. To the best of our knowledge, there has been no report on the electrical conductivity of the phosphorene, so in this work, when fitting the relaxation time of the bulk black-P, the anisotropy will be considered and the two in-plane values will be used in the case of phosphorene. We use the experimental electrical conductivity of bulk black-P reported in Ref. 19 and the fitted relaxation times $\tau$ at the temperature 300 K along the *a*, *b* and *c* axes corresponding to the crystal structure are $1.34\times10^{-13}$ s, $2.82\times10^{-14}$ s and $6.20\times10^{-14}$ s, respectively. By inserting the fitted values of $\tau$, we plot the curve of the electrical conductivity $\sigma$ of bulk black-P as a function of the chemical potential $\mu$ at 300 K, as shown in Fig. 3(a). Within the rigid band model, $\mu$ determines the doping level of the system. For the *n*-type doping, the Fermi level moves up ($\mu>0$); for the *p*-type doping, the Fermi level moves down ($\mu<0$). The corresponding carrier concentrations can be obtained by integrating the density of states (DOS) of the system. We can see from Fig. 3(a) that the electrical conductivity of bulk black-P exhibits anisotropic behavior. The electrical conductivities along the two in-plane directions (*a* and *c* axes) are generally larger than that along the out-of-plane direction (*b* axis). The difference of the electrical conductivities along the two in-plane directions is relatively smaller, with the value along the *c* axis slightly larger than that along the *a* axis in the range of $-0.5\,\text{eV} < \mu < 0.5\,\text{eV}$.

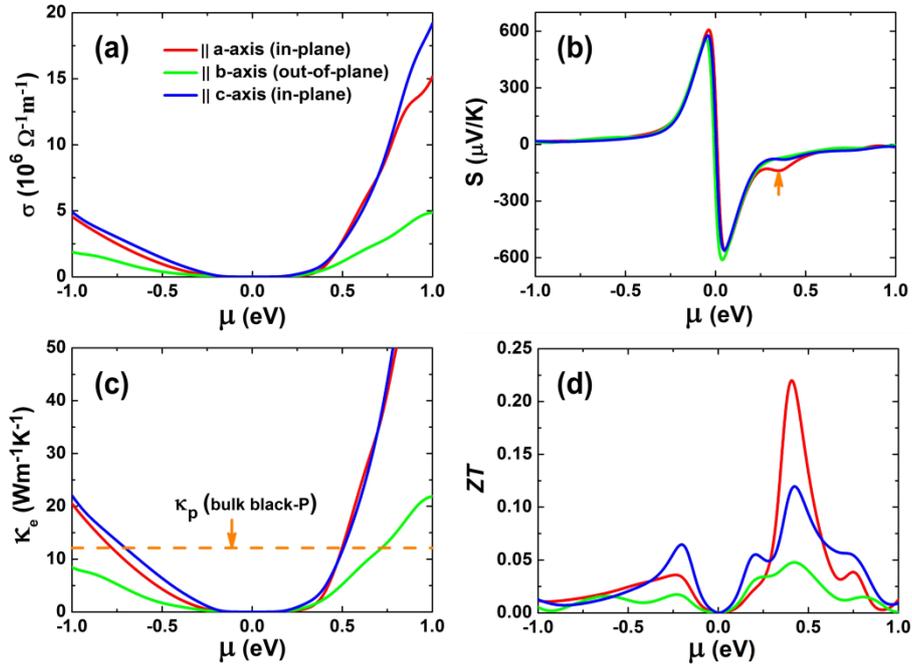

**Figure 3** Thermoelectric transport coefficients as a function of the chemical potential for bulk black-P: (a) electrical conductivity; (b) Seebeck coefficient; (c) electronic thermal conductivity and (d) $ZT$ value. The red, green and blue lines represent the coefficients along *a*, *b* and *c* axis, respectively.

Figure 3(b) shows the calculated Seebeck coefficient $S$ of bulk black-P as a function of the chemical potential $\mu$ at 300 K. In the vicinity of the Fermi level ($\mu = 0$), the Seebeck coefficient exists two obvious peaks for both *p*- and *n*-type doping. The absolute value of the Seebeck coefficient can reach as high as 612 μV/K at $\mu = 0.037$ eV, much larger than that of the conventional thermoelectric material $Bi_2Te_3$.[22] The difference of the Seebeck coefficient along the three directions is very small, except that there exists another small peak of −140 μV/K at $\mu = 0.35$ eV along the *a* axis, which is denoted by the arrow in Fig. 3(b). This local maximum may result in the peak of the $ZT$ value along this direction, which will be discussed later. In Fig. 3(c), we plot the electronic thermal conductivity $k_e$ as a function of the chemical potential $\mu$ at 300 K. We see that the overall topology of the electronic thermal conductivity $k_e$ is almost the same as that of the electrical conductivity $\sigma$, since $k_e$ is calculated from $\sigma$ based on the Wiedemann–Franz law $k_e = L\sigma T$. The dashed line in the figure denotes the experimental value of lattice thermal conductivity of bulk black-P (12.1 Wm$^{-1}$K$^{-1}$).[13] In the range of $-0.5\,\text{eV} < \mu < 0.5\,\text{eV}$, the electronic thermal conductivity is generally smaller than the lattice thermal conductivity. For the *n*-type doping, at the doping level of $\mu = 0.5$ eV, the electronic and lattice thermal conductivity become comparable. Moreover, the out-of-plane electronic thermal conductivity is relatively smaller than the in-plane ones. This is reasonable since there will be more scattering in the out-of-plane direction. Based on the calculated electronic transport coefficients and by using the lattice thermal conductivity reported experimentally,[13] we are now able to evaluate the $ZT$ value of the bulk black-P at room temperature, which is plotted in Fig. 3(d) as a function of the chemical potential $\mu$. Generally, the $ZT$ value at $\mu > 0$ is much larger than that at $\mu<0$, indicating that for the bulk black-P, the *n*-type doping is more favorable for the thermoelectric application than the *p*-type doping. On the other hand, the $ZT$ value exhibits the property of anisotropy, that is, the values in the two in-plane directions are much larger than that along the out-of-plane direction. For the in-plane directions, the $ZT$ value along the *c* axis is relatively larger than that along the *a* axis for the *p*-type doping, however, the case is reversed when the doping is *n*-type, where the $ZT$ value can reach as high as 0.22 at $\mu=0.41$ eV along the *a* axis. Note that at the chemical potential where the maximum of the two in-plane $ZT$ values appears, both the electrical conductivity and electronic thermal conductivity along the two directions differ little from each other. This is reminiscent of the local maximum of the Seebeck coefficient along the *a* axis (denoted by the arrow in Fig. 3(b)) which we mentioned above. We see that the location of the maximum $ZT$ value is very close to that of the small peak of the Seebeck coefficient denoted by the arrow, so the largest $ZT$ value along the *a* axis is likely caused by the local maximum of the Seebeck coefficient in this direction. The largest $ZT$ values which can be achieved in bulk black-P by appropriate *p*- and *n*-type doping are summarized in Table II, with the corresponding chemical potential $\mu$, the carrier concentration *n*, the electronic transport coefficients ($S$, $\sigma$, $\kappa_e$ and the thermoelectric power factor $PF=S^2\sigma$) and the directions. From Table II, we can see that there must be a trade-off between the Seebeck coefficient and electrical conductivity to get a largest *PF* and thus a largest $ZT$ value. Although the largest Seebeck coefficient of −612 μV/K is predicted at $\mu=0.037$ eV in this system, at that doping level, the electrical conductivity is not large enough to get a comparable *PF*. At the optimal doping level where the largest $ZT$ value of 0.22 is reached, the Seebeck coefficient $S$ is −114.6 μV/K and electrical conductivity $\sigma$ is 9.02×10$^5$ $\Omega^{-1}m^{-1}$, which result in the *PF* of 118.4 μWcm$^{-1}$K$^{-2}$, even larger than those outstanding thermoelectric materials reported experimentally.[2,3,5] However, due to the relatively large thermal conductivity ($\kappa_p$=12.1 Wm$^{-1}$K$^{-1}$, $\kappa_e$=4.06 Wm$^{-1}$K$^{-1}$), the largest $ZT$ value is 0.22 at 300 K. If the lattice thermal conductivity $\kappa_p$ can be reduced by intercalating scattering units into the inter-layer sites, the thermoelectric performance of bulk black-P could be further enhanced.

**Table II** Maximum $ZT$ values of bulk black-P and phosphorene for both *p*- and *n*-type doping. The corresponding chemical potential $\mu$, the carrier concentration *n* (in units of $10^{20}$ cm$^{-3}$ and $10^{12}$ cm$^{-2}$ for bulk black-P and phosphorene, respectively), the electronic transport coefficients ($S$, $\sigma$, PF and $\kappa_e$) and the directions are also shown. Note here we use the carrier concentration per area for phosphorene.

|   |   | $\mu$ (eV) | *n* ($10^{20}$ cm$^{-3}$/ $10^{12}$ cm$^{-2}$) | $S$ (μV/K) | $\sigma$ ($10^5$ Ω$^{-1}$m$^{-1}$) | PF (μW cm$^{-1}$K$^{-2}$) | $\kappa_e$ (Wm$^{-1}$K$^{-1}$) | ZT | Direction |
|---|---|---|---|---|---|---|---|---|---|
| Bulk black-P | *p*-type | −0.20 | 0.099 | 159.1 | 1.07 | 27.1 | 0.48 | 0.06 | *c* axis |
|   | *n*-type | 0.41 | 1.93 | −114.6 | 9.02 | 118.4 | 4.06 | 0.22 | *a* axis |
| Phosphorene | *p*-type | −0.90 | 6.04 | 198.7 | 3.52 | 138.9 | 1.58 | 0.30 | *c* axis |
|   | *n*-type | 0.90 | 4.64 | −189.4 | 2.68 | 96.3 | 1.21 | 0.22 | *c* axis |

The promising thermoelectric performance of bulk black-P makes us expect that the phosphorene may perform better since it was theoretically predicted that low-dimensional structures could have much larger $ZT$ values than their bulk counterparts.[23,24] As mentioned above, we use the in-plane relaxation times fitted for the bulk black-P when dealing with the electrical conductivity of the phosphorene. The calculated electrical conductivity $\sigma$ as a function of the chemical potential $\mu$ at 300 K is plotted in Fig. 4(a). The same as the case of

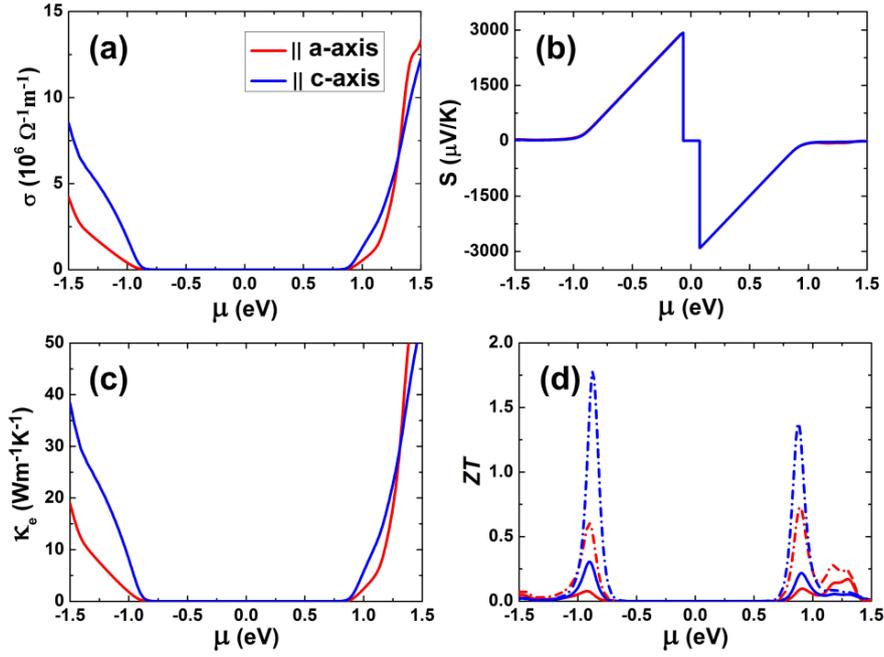

**Figure 4** Thermoelectric transport coefficients as a function of the chemical potential for phosphorene: (a) electrical conductivity; (b) Seebeck coefficient; (c) electronic thermal conductivity and (d) $ZT$ value. The dashed lines indicate the $ZT$ values calculated using lattice thermal conductivity of 1.21 Wm$^{-1}$K$^{-1}$. The red and blue lines represent the coefficients along $a$ and $c$ axis, respectively.

bulk black-P, the electrical conductivity along the $c$ axis is relatively larger than that along the $a$ axis. In the range of $-1.0\,\text{eV} < \mu < 1.0\,\text{eV}$ and along the $a$ axis direction, the electrical conductivity for the $n$-type doping is larger than that for the $p$-type doping at the same doping level (i.e., the absolute value of the chemical potential is equal to each other). The case of the electrical conductivity along the $c$ axis is just reversed. The Seebeck coefficient as a function of the chemical potential at 300 K is shown in Fig. 4(b). The two lines for the two different directions coincide with each other, indicating that the Seebeck coefficient of the phosphorene is isotropic. The maximum Seebeck coefficient can reach as high as 2917 μV/K at $\mu = -0.067$ eV, which is very beneficial for the thermoelectric applications. Figure 4(c) plots the electronic thermal conductivity as a function of the chemical potential at 300 K for phosphorene. In the range of $-1.0\,\text{eV} < \mu < 1.0\,\text{eV}$, the electronic thermal conductivity along the $a$ axis is below 2.42

Wm$^{-1}$K$^{-1}$ and the value is below 8.25 Wm$^{-1}$K$^{-1}$ along the *c* axis. We estimate the $ZT$ value of phosphorene at 300 K by using the experimental lattice thermal conductivity of bulk black-P (12.1 Wm$^{-1}$K$^{-1}$), which is plotted as a function of chemical potential in Fig. 4(d) (the solid lines). Although the electronic thermal conductivity along the *c* axis is larger than that along the *a* axis, the $ZT$ value along the *c* axis is much larger than that along the *a* axis. When comparing the electronic transport coefficients ($S$, $\sigma$ and $\kappa_e$), we can find that the anisotropy of the electrical conductivity dominates in determining the anisotropic property of the $ZT$ value. The largest $ZT$ value of 0.30 can be reached in the direction of *c* axis when the system is appropriately *p*-type doped ($\mu = -0.90$ eV). The optimal doping concentrations (here we use the carrier concentration per area) as well as the electronic transport coefficients with respect to the maximum $ZT$ value for both the *n*- and *p*-type doped systems are summarized in Table II. We can see that compared with the bulk black-P, larger *PF* of 138.9 μWcm$^{-1}$K$^{-2}$ can be achieved in phosphorene, with the Seebeck coefficient of 198.7 μV/K and electrical conductivity of 3.52×10$^5$ Ω$^{-1}$m$^{-1}$. Note here the lattice thermal conductivity of bulk black-P is used for the estimation of the $ZT$ value. If the lattice thermal conductivity of the phosphorene can be reduced by an order of magnitude (that is, 1.21 Wm$^{-1}$K$^{-1}$), which can be realized in many low-dimensional structures, the $ZT$ value of the phosphorene can be enhanced up to 1.78. The corresponding $ZT$ value as a function of chemical potential by inserting the lattice thermal conductivity of 1.21 Wm$^{-1}$K$^{-1}$ is plotted in Fig. 4(d) by the dashed lines. The large $ZT$ value of the phosphorene indicates that this system is a promising thermoelectric material.

In summary, we have investigated the electronic and thermoelectric properties of bulk black-P and the corresponding phosphorene. Both of them are semiconductors with direct band gaps of 0.34 eV and 1.80 eV, respectively. The electronic transport coefficients are calculated using semi-classical Boltzmann theory based on the electronic structure. For the bulk black-P, at the optimal doping level, a large *PF* of 118.4 μWcm$^{-1}$K$^{-2}$ can be achieved. However, due to the relatively higher lattice thermal conductivity (12.1 Wm$^{-1}$K$^{-1}$), the largest $ZT$ value is 0.22 in bulk black-P. If the lattice thermal conductivity can be reduced by intercalating scattering units into the inter-layer sites, the thermoelectric performance of bulk black-P could be further enhanced. For the phosphorene, even larger *PF* of 138.9 μWcm$^{-1}$K$^{-2}$ can be achieved by properly *p*-type doping. The $ZT$ value can reach 0.30 when conservatively estimated by using the bulk lattice thermal conductivity. If the lattice thermal conductivity could be reduced by an order of magnitude compared with the bulk black-P, a $ZT$ value as high as 1.78 could be obtained in the phosphorene. Our calculations indicate that both the bulk black-P and the phosphorene are very promising thermoelectric materials, which deserve more investigations in experiment.

**Computational Methods**

The structural and electronic properties of black-P are investigated using the first-principles pseudopotential method as implemented in the ABINIT code[25,26,27]. The Brillouin zones are sampled with 8×8×10 and 12×1×10 Monkhorst–Pack *k*-meshes for the bulk black-P and phosphorene, respectively. The cutoff energy for the plane wave expansion is set to be 800 eV. For the structural relaxation, the exchange-correlation energy is in the form of Perdew–Burke–Ernzerhof (PBE)[28] with generalized gradient approximation (GGA); for bulk black-P, the van der Waals interactions are treated by the vdW-DFT-D2 functional.[29] Both the geometries and atomic positions are fully relaxed until the force acting on each atom is less than 0.5×10$^{-3}$ eV/Å. In the calculations of the electronic structures, we use the TB-mBJ exchange potential[30] in combination with the correlation potential of PW91,[31] which can reproduce accurate band gaps for many semiconductors. Based on the electronic structure, the electronic transport coefficients are derived by using the semi-classical Boltzmann theory within the relaxation time approximation[32] and doping is treated by the rigid band model.[33] The electronic thermal conductivity $\kappa_e$ is calculated using the Wiedemann–Franz law $k_e = L\sigma T$, where $L$ is the Lorenz number. In this work, we use the Lorenz number of 1.5×10$^{-8}$ WΩ/K$^2$ for the non-degenerate semiconductor.[1]

**Corresponding Author**

* Email: wjlu@issp.ac.cn. * Email: ypsun@issp.ac.cn

**Acknowledgments**

This work was supported by the National Key Basic Research under Contract No.2011CBA00111, the National Nature Science Foundation of China under Contract No.11274311, the Joint Funds of the National Natural Science Foundation of China and the Chinese Academy of Sciences' Large-scale Scientific Facility (Grand No.U1232139), and Anhui Provincial Natural Science Foundation under Contract No.1408085MA11. The calculation was partially performed at the Center for Computational Science, CASHIPS.